\documentclass[useAMS,usenatbib]{mn2e}

\usepackage{epsfig}

\def\lesssim{\mathrel{\hbox{\rlap{\hbox{\lower4pt\hbox{$\sim$}}}\hbox{$<$}}}}
\def\gtrsim{\mathrel{\hbox{\rlap{\hbox{\lower4pt\hbox{$\sim$}}}\hbox{$>$}}}}

\title[Multi-Planet Systems with Debris Disks]
{The Total Number of Giant Planets in Debris Disks with Central Clearings}

\author[Faber \& Quillen]
{
Peter Faber  \& Alice C. Quillen  \\
Department of Physics and Astronomy, University of Rochester,
Rochester, NY 14627;
pfaber@mail.rochester.edu; aquillen@pas.rochester.edu}

\begin{document}
\label{firstpage}
\maketitle
\begin{abstract}
Infrared spectra from the Spitzer Space Telescope  (SSC)
of many debris disks are well fit with a single black body temperature
which suggest clearings within the disk.
We assume that inside the clearing
orbital instability due to planets removes dust generating
planetesimal belts and dust generated by
the outer disk that is scattered or drifts into the clearing.
From numerical integrations we estimate a minimum planet
spacing required for orbital instability (and so planetesimal
and dust removal) as a function of system age and planet mass.
We estimate that a $10^8$ year old debris disk with a dust disk edge
at a radius of 50 AU hosted by an A star must contain
approximately 5 Neptune mass planets between
the clearing radius and the iceline in order to 
remove all primordial objects within it.  
We infer that known debris disk systems contain at
least a fifth of a Jupiter mass in massive planets.
The number of planets and spacing
required is insensitive to the assumed planet mass.
However an order of magnitude higher total mass in planets could reside
in these systems if the planets are more massive. 
\end{abstract}

\section{Introduction}

Recent observations made with the Spitzer Space Telescope (SSC) have
added to the total number of known debris and circumstellar disks
\citep{rieke05,beichman05,gorlova06,chen06,bryden06,beichman06}.
Theses disks have total opacity, as estimated from the fraction
of stellar light re-emitted in the infrared, in the range
10$^{-3}$ - 10$^{-5}$.  
The fraction of stars with disks increases
from 2\% for $\tau > 10^{-4}$ to 12\% for $\tau > 10^{-5}$ \citep{bryden06},
and the fraction of stars with disks detected increases with decreasing
stellar age \citep{rieke05,gorlova06}.  Infrared spectra of many debris
disks are well fit with a single black body temperature
suggesting that they possess inner holes \citep{chen06}.
Planets are expected to be responsible for the inner clearings, however
the number and masses of these planets has yet to be constrained
by dynamical arguments. 
An examples of a nearby system with an inner clearing
that has been resolved with imaging is  
the Fomalhaut system (e.g., \citealt{kalas05}).

With the discovery of multiple planet extrasolar systems,
the stability of planetary systems has been investigated with renewed
interest (e.g., \citealt{ford01,barnes04,lepage04,ford05,raymond05,barnes06,chatter07}).
A system of two planets on low-inclination and eccentricity
orbits will never experience mutual close encounters if the initial semi-major axis
difference, $\Delta_m$, measured in mutual Hill radii, exceeds $2 \sqrt{3}$ 
\citep{gladman93}. 
For multi-planet systems, \citet{chambers96} numerically investigated
the dependence of the timescale of 
the first planetary encounter, $t_e$, on the planet mass, $M_p$, number of 
planets, $N_p$, and the spacing between them, $\Delta_m$.
When all planets have the same mass, 
they found that $\log_{10}$ of the timescale 
is proportional to the planet spacing in Hill radii. 
When the number of planets $N_p \gtrsim 5$, the timescale is not strongly
dependent on the number of planets.
Here $\mu = M_p/M_*$ is the planet mass in units of the stellar mass.
The total mass in planets in a system would be
$M_{p,total} = M_p N_p$.

For the number of planets $N_p=3$,  the time to first encounter, $t_e$,
varies from $10^4$ to $10^6$ years for planet masses
$\mu$ ranging from $10^{-5} - 10^{-9}$ where a year represents the orbital
period of the innermost planet.  
Large variations in planet mass are required to significantly change
the timescale required for instability to develop
(e.g., see Figure 4 by \citealt{chambers96}).
The time to first planet/planet encounter is much more sensitive 
to planetary spacing than it is to planet mass
since the hill radius is only dependent on mass ratio to the one third.
This implies that a higher total mass in planets
can be resident in a stable system when the mean
planet mass is higher even though the planets 
must be somewhat further apart.
A nearly unstable system with a larger number of Earth mass planets must
have spacings such that the system has lower total mass in planets 
than a nearly unstable system containing
small number of somewhat more widely spaced Jupiter mass planets. 
For example, a planetary system that has not suffered a near encounter in
$10^6$ years has spacing greater than distance
in mutual Hill radii, $\Delta_m >7.5$ for planet mass ratio $\mu =10^{-9}$
but only $\Delta_m > 6$ for $\mu =10^{-5}$ and .

For a system observed with a dusty disk containing an interior clearing,
we assume 
that bodies residing between the planets must have been removed via
dynamical processes on a timescale shorter than the age of the system.
If numerous planetesimals lie interior to the clearing
they could efficiently produce dust in the clearing.
Dust produced in the outer disk can scatter
into the clearing because the 
collision timescales estimated from the dust opacity
is significantly shorter than the ages of the systems.  
Dust can also drift into
the clearing because of radiative forces. 
Studies of dust particle integrations suggest that
massive planets,  at least Neptune in size, 
are required to account for a clearing in the dust distribution 
(e.g., \citealt{ozernoy00,liou99,moromar02,deller05,quillen06,quillen07}).
Sharp eccentric edges such as that in the Fomalhaut system
\citep{kalas05} imply that clearing
by planets is required in some systems \citep{quillen06}. 
Our assumption is consistent with scenarios for the evolution of our solar system.
Most orbits in between Jupiter and Neptune become planet orbit crossing
within the age of the solar system \citep{duncan89}.
Nevertheless cratering history suggests that there
was an epoch of early bombardment possibly associated
with the clearing of interplanetary debris 
(e.g., \citealt{gomes05,bottke05,strom05}).
We note that some systems with detected
IR excesses may contain evidence for inner exo-zodiacal belts 
(e.g., HD12039; \citealt{hines06}) in their
spectral energy distribution or they may be present but
below the detection limit. Nevertheless if planetesimals are originally 
widely distributed and contain enough mass to form planets (as
suggested by the cratering record in our solar system),
then the lack of bright or detected  exo-zodiacal belts implies that
most planetesimals have been ejected, accreted onto planetary bodies or
have fallen into the central star.

In this paper we combine the assumptions that planetesimals
within a dust disk edge must be removed within the age of
the system with estimates of the timescale $t_e$
for the first planet/planet encounter in a multiple planet system.
While other explanations for clearing exist, such as a single planet
migrating outward constantly eroding away the 
inner edge of the disk \citep{gomes05},
we only consider clearing via multiple planet induced instability.
We estimate a minimum spacing between planets such that
objects placed between them are likely to be 
removed on a timescale shorter than the age of the system
while the planets themselves remain stable.
For simulations with planets, only
the timescale until first planet/planet encounter depends
on the planetary spacing and planet mass.    
Here we estimate the planet spacing for the instability of
inter planetary objects by requiring the planets to be 
twice as far away as required for a first planet/planet
encounter to occur during the age of the system.
Such a system will have planets which will be stable on a timescale much
longer than the age of that system. 
However, particles between the planets will have timescales
shorter than the age of the system (e.g., \citealt{duncan89}).
We expect that interplanetary particles residing in the system
should have instability timescales similar to
those estimated from N-body simulations of planets spaced at half
the spacing.   By doubling the spacing, we ensure that under these
timescales, the planets will be stable while the planetesimals
will be cleared.  The age of the system is a
timescale for planetesimal clearing, not planetary instability.
As all massive bodies influence each other
and low mass bodies do not influence massive ones, we expect
that our estimate for the required planetary spacing is
conservative.   Our spacing estimate will be larger
than the actual one required for efficient clearing of
interplanetary bodies and so will lead to a lower
limit on the number of, and total mass, in planets required to clear
particles from clearings. 

\section{Numerical Integrations}

Integrations were done using John Chambers'
MERCURY package version 6. 
\footnote{Available to download at: http://www.arm.ac.uk/$\sim$jec/}
Mercury6 contains several N-body algorithms.
We use the hybrid symplectic/Burlisch integrator since it is 
relatively fast and has the ability to compute close encounters.

Ten massive bodies of all the same
mass ratio were used.  \citet{chambers96} finds that 
as long as the number of planets
in a system exceeds five, then the mean stability is 
nearly independent of number of planets.
That is, a
system of five planets has similar close encounter timescale dependence
on separation as a system of ten or twenty planets.  
As we desire an estimate for the number of planets,
we compute the stability timescale for a larger number of bodies in
such a way as to be insensitive to actual number of planets.
Each body has zero initial eccentricity, inclination,
longitude, periapse and mean anomaly.  All planets were assumed
to have density 1 g~cm$^{-3}$.  They felt no
non-gravitational forces.    

The initial semi-major axis, $a_{n+1}$, 
of the $n+1$-th planet was initially set to depend
only on the semi-major axis of the planet just interior or $a_n$
with a spacing $\delta$ such that 
\begin{eqnarray}
a_{n+1} & =   & \left(1 + \delta \right) a_n   \\
        & =   & \left[1+ \Delta \left({r_H \over a_n}\right) \right] a_n 
\nonumber  \\
& = & \left(1+ \Delta \sqrt[3]{\mu \over 3} \right) a_n, 
\nonumber
\label{eqn:delta}
\end{eqnarray}
where $r_H$ is the Hill radius for the $n$-th planet
and $\Delta$ is the spacing in Hill radii. 
The spacings chosen for our integrations 
ranged from $\Delta = 2.5$ to 11.0.  
This definition for planet spacing is somewhat different than the 
spacing used by \citet{chambers96}.  They use multiples of the mutual
Hill radius rather than the Hill Radius.  For small values of
$\mu$, the definitions are the same except for a factor of $2^{1/3}$.

We carried out a series of integrations for each planet mass
ratio with different initial values of $\delta$ setting
the planetary spacings.
Timescales for the simulation are given in orbital
periods of the innermost planet.
Integrations were calculated for ten million years or until
a close encounter occurred.  A close encounter is defined as when the
distance between two planets was less than one Hill radius. 
When this circumstance occurred, the system was assumed
unstable and the integration terminated.  
The integrations did not include collisions.
After every 1000 years,
the six orbital elements for each planet was output
and recorded.  Planets were ejected from the system
if their semi-major axis exceeded 200 AU. Mass ratios of 
$10^{-3}$, $10^{-5}$, and $10^{-7}$ were used. 
We have used larger mass ratios than \citet{chambers96}
so we can consider systems with massive planets.

\begin{figure}
\includegraphics[angle=0,width=3.6in]{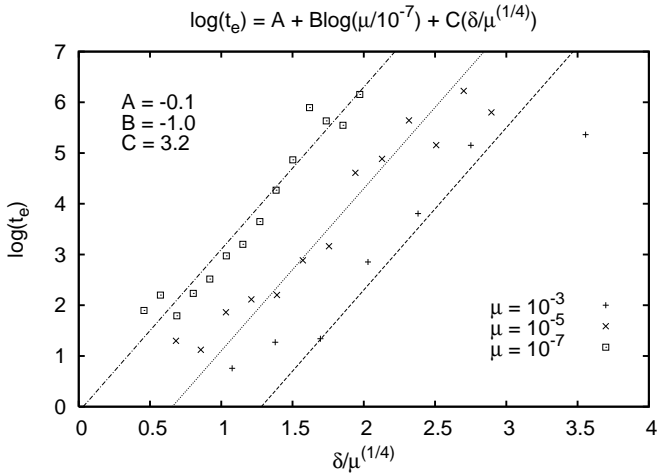}
\caption
{
\label{fig:dt}
The relationship between planet spacing, $\delta$,
(see equation \ref{eqn:delta})
and time to first close planet/planet encounter,  
for systems with 10 coplanar planets.
The spacings are multiplied by $\mu^{1/4}$ so that
the sets of integrations with the same planet mass
lie on lines with the same slope.
Three sets of integrations are shown, those with
planet mass ratio $\mu=10^{-3}$ (pluses), $10^{-5}$(x's) 
and $10^{-7}$ (open squares).
A fit to the three sets of integrations is shown
with the solid lines.
The fit is described with a single
equation that relates timescale of first encounter
to planet mass ratio and spacing (equation \ref{relationship}).
}
\end{figure}

The timescale to first close encounter time for our integrations are 
plotted as a function of planet spacing $\delta/\mu^{1/4}$ 
in Figure \ref{fig:dt}.  \citet{chambers96} found a linear
relationship between timescale of first encounter
and planetary spacing in Hill radii but the slope of this relation
is dependent on planet mass ratio.  Only when rescaled by $\mu^{1/4}$
(see their figure 4) does the slope become independent
of mass ratio. We confirm this here.

We have fit the points shown in Figure \ref{fig:dt} 
measured from our integrations with a single
equation;
\begin{equation}
\log_{10}({t_e \over yrs}) = -0.1 - \log_{10}(\mu/10^{-7}) + 3.2(\delta/\mu^{1/4}).
\label{relationship}
\end{equation}
This relationship is shown with solid lines
for the three mass ratios shown in Figure \ref{fig:dt}.
The slopes of the lines in Figure \ref{fig:dt} are consistent
with those found by \citet{chambers96}.

The above equation is useful as it can be used
to estimate a minimum spacing required for instability
as a function of system age.
Rearranging equation \ref{relationship} we find   
\begin{equation}
\delta = {\mu^{1/4} \over 3.2} \left[ \log_{10}({t_e \over yrs})
+ 0.1 +\log_{10}(\mu / 10^{-7}) \right].
\label{key}
\end{equation} 
We confirm that $\delta$ is not strongly dependent on planet mass,
as we discussed in Section 1.  Faults in this equation are 
discussed in the final section of this paper.

\section{Estimate of a lower limit on total mass in planets} 

We use the relationship measured from the integrations
(equation \ref{key}) to
estimate the minimum spacing between planets required
for bodies to be removed within a clearing.
We set $t_e$ in the above equation to the age of the system,
requiring particle removal in the clearing within the lifetime of the star.
Assuming that all planets have the same mass ratio,
and that they are all spaced at twice $\delta$  
(defined by equation \ref{eqn:delta}),
the total number of planets, $N_p$, must be larger 
than that satisfying
\begin{equation}
(1+2\delta)^{N_p} = {R_{out} \over R_{in}},
\label{eqn:Np}
\end{equation}
where $R_{out}$ is the semi-major axis of the outermost
planet and $R_{in}$ is the semi-major
axis of the innermost planet.
The factor of two is used to ensure that over this
timescale, the planetesimals are cleared while
the planets themselves remain stable.
Equation \ref{eqn:Np} for $N_p$ can be solved given
$R_{in}, R_{out}$,  and as $\delta$ depends on $t_e$ and $\mu$,
given the age of the star and an assumed planet mass ratio.

The outermost planet must reside within the dust disk clearing
so we can set $R_{out}$ to the radius corresponding to
the dust temperature estimated from infrared spectra.
We conservatively estimate $R_{in}$ to be
the distance at which water freezes (the iceline) based on the 
luminosity of the central star and given by:
\begin{equation}
R_{in} = \left[ L_* \over 4 \pi \sigma T_f^4 \right]^{(1/2)},
\label{eqn:iceline}
\end{equation}   
where $T_f = 273$K is the freezing point of water, $L_*$
is the luminosity of the star, and $\sigma$
is the Stefan-Boltzmann constant.
The iceline represents a convenient limit for the inner radius.
While massive planet formation models favor formation
outside the iceline, planets could migrate closer to the star.
Massive planets could reside within the iceline
so our estimate for the total number of planets
is a lower limit.

The number of planets in a system is then given by:
\begin{equation}
N_p = {\log_{10}(R_{out}/R_{in}) \over \log_{10}(1 +  2\delta)}.
\label{eqn:Npcalc}
\end{equation}
The factor of 2 for $\delta$ on the right hand
side follows because
we have set the planets to be twice as far away as that required
for instability in the pure 10-body system (see discussion in section 1).

The remaining quantity that we must choose  
to estimate the number of planets is the
planet mass ratio, $\mu$.
Recent studies suggest that maintenance of a clearing
in the dust distribution requires at least a Neptune
mass planet 
(e.g., \citealt{ozernoy00,liou99,moromar02,deller05,quillen06,quillen07}).
We adopt this mass ratio, $\mu = 5 \times 10^{-5}$ as a starting point
and then will consider the effect of varying this ratio.
Once we estimate the number of planets present, the
total mass in planets 
\begin{equation}
M_{p,total} \ga \mu M_* N_p.
\end{equation} 
The actual mass in planets is likely to be higher than the above estimate
for a number of reasons:
There may be massive planets within the iceline.
We have assumed that all planets have the same mass.
If more massive planets are present then a larger
total planet mass could be present.  This follows
because $\delta$ is not strongly dependent on planet mass.
Our integrations were run with ten planets but we may estimate
that the number of required planets is fewer and systems
with fewer planets are likely to be more stable.

We apply our framework for estimating the minimum number and
total mass in planets to the disks studied
by \citet{chen06}.  \citet{chen06} has measured
the radii of the clearings in the dust based on dust temperatures
measured from spectra observed with the Infrared Spectrograph
on board the SSC.  Their sample consists of 
main sequence stars which are located within 150pc and   
stars that are part of OB associations or moving groups.
The properties of these stars are listed in Table \ref{tab:stars}
with our calculated limits on the
number of planets, $N_p$, and total mass in planets, $M_{p,total}$ 
within the clearings.
We find that 3-6 planets of Neptune mass ratio are required
and the total mass in planets is likely to be larger than $\sim 0.2 M_J$
where $M_J$ is a Jupiter mass.

\begin{figure}
\includegraphics[angle=0,width=3.6in]{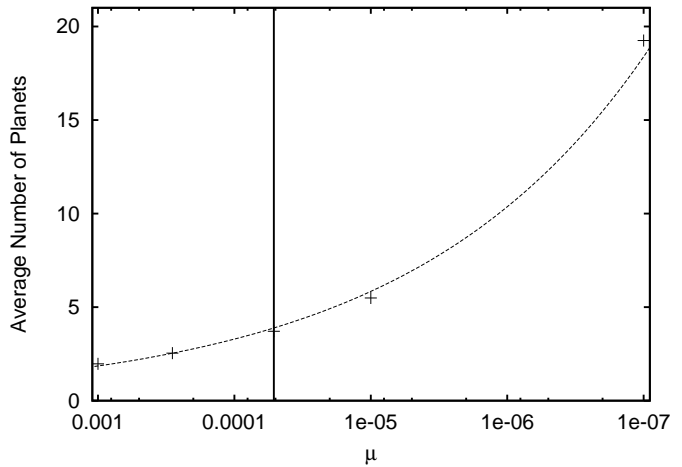}
\label{uplan}
\caption
{
\label{fig:chart}
We consider the average number of planets for the systems in 
Table \ref{tab:stars} for assumed mass ratios between
$10^{-3}$ and $10^{-7}$. The curved line, $\propto
(\log_{10}\mu)^{-0.25}$,
is a best fit line for the average number of planets in a system based on
mass ratio.   We find that the number of estimated
planets is insensitive to the assumed planet mass ratio
for planets sufficiently massive to effectively
empty a clearing.  If more massive planets are present
then the total mass in planets could be an order of
magnitude higher than listed in Table \ref{tab:stars}.
}
\end{figure}

Our estimates for the number of planets and total mass in
planets depends on the assumed planet mass ratio, $\mu$.
We investigate how the estimates depend on the assumed ratio.
From the same stellar sample we have computed the mean number of planets
using $\mu$ ranging from $10^{-3}$ to $10^{-7}$.  
The results are plotted in Figure \ref{fig:chart}.
The vertical line on this plot shows
the location for Neptune's mass ratio. 
We find that the total number of planets
is insensitive to the assumed mass ratio for $\mu > 10^{-5}$.
Planets with ratios lower than this  are unlikely to be
able to effectively eject material from central clearings within
the lifetime of the system.
While the total number of estimated planets is likely to be
low, the total mass in planets could be an order of magnitude
higher than given in Table 1. Jupiter mass planets
with similar number and only slightly larger spacings would be
capable ejecting material from the clearings during the lifetime
of these systems.

It is interesting to compute the total number of planets 
for our solar system assuming a dust belt in the Kuiper Belt
were detected (as predicted by \citealt{liou99}).
Using $R_{out} = 30$ AU and and an iceline at 1.0 AU for $R_{in}$
we would predict $N_p = 6.6$ Neptune mass planets,
giving a total planetary mass of 0.36 $M_J$.
Much lower than the actual planetary mass in our solar system
in this semi-major axis range (1.4 $M_J$), we reiterate that our
caclulations represent
an absolute lower limit on the total planetary mass of a system.
Our solar system easily increases its planetary
mass since the planets closer than Neptune have varying mass ratios
larger than that of Neptune.  Since stabilty is less dependent
on mass ratio than spacing, fewer more massive planets exist, 
thus increasing total planetary mass in the solar system. 
  
\section{Conclusions and Discussion}

In this paper we have made the assumption that bodies,
both planetesimal and dust particles,
within a dust disk clearing must be removed from 
the clearing within the age of
the system via gravitational interactions with planets.  
This assumption is consistent with some scenarios
for the early evolution of our solar system.
This assumption is combined
with a rough estimate for the timescale
for instability in multiple planet systems to estimate
the planet spacing required for instability. 
The result is an estimate for the number of planets likely
to reside in debris disk systems that host disks with clearings.
We find that the number of required planets is 3-6, between
the dust disk clearing radius and the iceline, for
the sample of disks studied by  \citep{chen06}, assuming
a planet mass ratio like that of Neptune.  
We find that the number
of estimated planets is only
weakly dependent on the assumed planet mass.  At least
0.2 $M_J$ total mass in planets is required in each system.
The planets could be more massive and an order of magnitude
more mass in planets could reside in these systems.

Our estimate neglects planets that lie within
the iceline but planets could migrate or form closer to the star.
The estimate for the planet spacing is based
on integration of a 10-body system but our estimated planet
number falls below 10.  
Further numerical experiments show that if the number of planets
is decreased to five, a decrease in $\delta / \mu^{1/4}$
by ten percent results in the same instability timescale.
The estimate for spacing is twice that for 
instability in a 10-body system but really
we require an instability timescale for massless bodies in
between massive bodies (e.g., as studied by \citealt{duncan89}).
We have assumed that the planet/planet encounter timescale
is related to a clearing timescale and this is not necessarily
true as small, low mass objects do not feel forces
on each other the way that large, massive bodies do.
A more sophisticated treatment would require
integration of a much larger number of particles and low mass 
as well as planetary mass bodies.  We could also examine
the effects of resonances on stability timescales and
the phase space of systems with massive planets as was done by
\citet{chatter07}.
Such an exploration could 
improve on the crude constraints on multi-planetary systems based on observations
of debris disks that we have explored here.
Here we have assumed that central clearing of debris and dust is primarily
due to scattering by planets. However collisional cascade
models for dust production in debris disks predict that the dust
production rate drops with time and that the centers of
disks evolve faster than than their outer parts (e.g., \citealt{DD03}). 
It is possible that future observations and statistical studies
may differentiate between collisional cascade evolution
and planetary clearing models.

We thank Richard Edgar, Eric Ford, and Alessandro Morbidelli
for interesting discussions and correspondence.
Support for this work was in part provided by National Science
Foundation grants AST-0406823 $\&$ PHY-0552695, the National
Aeronautics and Space Administration under Grant No.$\sim$NNG04GM12G
issued through the Origins of Solar Systems Program, and
HST-AR-10972 to the Space Telescope Science Institute.  

\begin{table*}
\begin{minipage}{120mm}
\caption{Stars with Dusty Disks with Clearings \label{tab:stars}}
\begin{tabular}{@{}lccccccc}
\hline
Name    & $M_*$ & $R_{out}$ & $R_{in}$ & $L_*$ & Age & $N_p$ & $M_{p,total}$ \\
        &($M_\odot$)& (AU) & (AU) & ($L_\odot$)   & (Gyr) &     &($J_\odot$) \\
\hline
$\gamma$ Cas & 4.0 & 72  & 16.5 & 250 & 0.11  & 3.2  &  0.17 \\
HR 333       & 2.5 & 34  & 6.67 & 41  & 0.11  & 3.5  &  0.19 \\
49 Cet       & 2.2 & 22  & 4.54 & 19  & 0.05  & 3.5  &  0.19 \\
HR 506       & 1.2 & 21  & 1.35 & 1.7 & 0.3   & 5.8  &  0.31 \\
$\gamma$ Tri & 2.6 & 31  & 7.65 & 54  & 0.17  & 3.0  &  0.16 \\
$\tau^3$ Eri & 2.0 & 12  & 3.75 & 13  & 0.5   & 2.4  &  0.13 \\
HR 1082      & 1.9 & 52  & 3.45 & 11  & 0.15  & 5.9  &  0.31 \\
HR 1570      & 2.2 & 42  & 4.88 & 22  & 0.23  & 4.6  &  0.24 \\
$\eta$ Lep   & 2.3 & 10  & 5.70 & 30  & 0.18  & 1.2  &  0.06 \\
HD 53143     & 0.8 & 4   & 0.72 & 0.48 & 0.97$^a$& 3.5 &  0.19 \\
HR 3314      & 2.5 & 27  & 6.83 & 43  & 0.19  & 2.9  &  0.16 \\
HD 95086     & 1.7 & 30  & 2.85 & 7.5  &0.016$^b$& 5.5 &  0.29 \\
$\beta$ UMa  & 2.7 & 53  & 9.32 & 80 & 0.34  & 3.6  & 0.19 \\
$\beta$ Leo  & 2.0 & 19  & 3.90 & 14 & 0.05  & 3.5  &  0.19 \\
HD 110058    & ?   & 20  & 3.29 & 10 & 0.016$^b$& 4.2 &  0.22 \\
$\gamma$ Boo & 2.2 & 23  & 4.77 & 21 & 0.27  & 3.3 & 0.18 \\
$\alpha$ CrB & 2.7 & 33  & 9.61 & 85 & 0.30  & 2.6 &  0.14 \\
HD 139664    & 2.7 & 15  & 1.97 & 3.6 & 0.48  & 4.2  & 0.23 \\
HD 146897    & 1.5 & 17  & 2.23 & 4.6 &0.005$^b$& 4.9  & 0.26 \\
HR 6297      & 1.8 & 21  & 3.45 & 11 & 0.42  & 3.8 &  0.20 \\
HR 6486      & 1.7 & 14  & 2.71 & 6.8 & 0.24  & 3.5 & 0.19 \\
HR 6532      & 3.0 & 47  & 10.05 & 93 & 0.2   & 3.3 & 0.17 \\
78 Her       & 2.6 & 23  & 7.44 & 51 & 0.05  & 2.5  & 0.13 \\
$\gamma$ Oph & 2.5 & 27  & 6.50 & 39 & 0.19  & 3.0 &  0.16 \\
HR 6670      & 1.5 & 13  & 2.11 & 4.1 & 1.6   & 3.6 &  0.19 \\
HD 181327    & 1.9 & 20  & 1.64 & 2.5 & 1.4   & 5.0 &  0.27 \\
HD 191089    & 1.8 & 14  & 1.83 & 3.1 & 1.6   & 4.1 &  0.22 \\
HR 8799      & 1.6 & 8   & 2.57 & 6.1 & 0.59  & 2.3 &  0.13 \\
SUN          & 1.0 & 30  & 1.04 & 1.0 & 4.6   & 6.6 &  0.36 \\
\hline
\end{tabular}
{
\\
The stellar masses ($M_*$), outer disk radii ($R_{out}$), Luminosity ($L_*$), 
and stellar ages are given for the sample discussed and studied by \citet{chen06}.
$^a$Age of HR 53143 from \citet{zuckerman04}.
$^b$Age of HR 95086, HD 146897  from \citet{deZeeuw99}.
Note as can be seen from Table 1 of \citet{chen06} there are 
discrepancies between available age estimates.
The rightmost two columns are calculated using 
Equations \ref{eqn:Npcalc} and \ref{key} and show
the estimated number of planets residing between the iceline
and the clearing edge (at $R_{out}$).
The iceline radius $R_{in}$ is estimated using equation 
\ref{eqn:iceline}.
Each planet is assumed to have the mass ratio of Neptune.
The number of planets, $N_p$, is estimated 
assuming that the planets are sufficiently close
together that all interplanetary debris has suffered a close
encounter with a planet within the lifetime of the system.
The total mass in planets 
$M_{p,total} = N_p \mu M_*$  is a lower limit for the
total mass in planets residing in the system.
Our own solar system is included to emphasize this point.
As is the case in our Solar System, if more massive
planets are present, then more
mass could be present in planets 
between the dust disk and the iceline even though the planets must
be further apart.
}
\end{minipage}
\end{table*}

{}


\begin{thebibliography}{}

\bibitem[Barnes \& Raymond(2004)]{barnes04}
Barnes, R., \& Raymond, S. N., 2004, ApJ, 617, 569

\bibitem[Barnes \& Greenberg(2006)]{barnes06}
Barnes, R., \& Greenberg, R.  2006, ApJ, 647, L163
	
\bibitem[Beichman et al.(2005)]{beichman05}
Beichman, C. A., et al. 2005, ApJ, 622, 1160 

\bibitem[Beichman et al.(2006)]{beichman06}
Beichman, C. A., et al  2006, ApJ, 652, 1674

\bibitem[Bottke et al.(2005)]{bottke05}
Bottke, William F., et al 2005, Icar, 179, 63B

\bibitem[Bryden et al.(2006)]{bryden06}
Bryden, G., Beichman, C. A.,  et al.  2006, ApJ, 636, 1098

\bibitem[Chambers et al.(1996)]{chambers96}
Chambers, J. E., Wetherill, G. W., \& Boss, A. P. 1996, Icarus, 119, 261

\bibitem[Chambers(1999)]{chambers99}
Chambers, J. E. 1999, MNRAS, 304, 793

\bibitem[Chatterjee et al.(2007)]{chatter07}
Chatterjee, Sourav, et al. 2007, arXic:astro-ph/0703166

\bibitem[Chen et al.(2006)]{chen06}
Chen, C. H.,  et al. 2006, ApJS, 166, 351


\bibitem[Deller \& Maddison(2005)]{deller05}
Deller, A. T., \& Maddison, S. T. 2005, ApJ, 625, 398

\bibitem[Dominik \& Decin(2003)]{DD03}
Dominik, C., \& Decin, G. 2003, ApJ, 598, 626

\bibitem[deZeeuw et al.(1999)]{deZeeuw99}
deZeeuw, P.T., Hoogerwerf, R., de Bruijne, J.H.J., 
Brown, A.G.A., \& Blaauw, A.  1999, AJ 117, 354

\bibitem[Duncan et al.(1989)]{duncan89}
Duncan, M., Quinn, T., \& Tremaine, S.  1989, Icarus, 82, 402

\bibitem[Ford et al.(2005)]{ford05}
Ford, E. B., Lystad, V., \& Rasio, F. A. 2005, Nature, 434, 873	

\bibitem[Ford et al.(2001)]{ford01}
Ford, E. B., Havlickova, M., \& Rasio, F. A. 2001, Icarus, 150, 303	

\bibitem[Gladman(1993)]{gladman93}
Gladman, B. 1993, Icarus, 106, 247

Origin of the cataclysmic Late Heavy Bombardment period of the terrestrial planets
\bibitem[Gomes et al.(2005)]{gomes05}
Gomes, R., Levison, H. F., Tsiganis, K., \& Morbidelli, A.	
2005, Nature, 435, 466

\bibitem[Gorlova et al.(2006)]{gorlova06}
Gorlova, N., Rieke, G. H., Muzerolle, J., Stauffer, J.R., 
Siegler, N., Young, E. T., \& Stansberry, J. H.	
2006, ApJ, 649, 1028

\bibitem[Hines et al.(2006)]{hines06}
Hines, D. et al.  2006, ApJ, 638, 1070

\bibitem[Kalas et al.(2005)]{kalas05}
Kalas, P., Graham, J. R., \& Clampin, M. 2005, Nature,  435, 1067

\bibitem[Lepage \& Duncan(2004)]{lepage04}
Lepage, I., \& Duncan, M. J. 2004, AJ, 127, 1755

\bibitem[Liou \& Zook(1999)]{liou99}
Liou, J. -C., \& Zook, H. A. 1999, AJ, 118, 580

\bibitem[Moro-Mart\`in \& Renu(2002)]{moromar02}
Moro-Mart\`in, A., \& Malhotra, R. 2002, AJ, 124, 2305M

\bibitem[Ozernoy et al.(2000)]{ozernoy00}
Ozernoy, L.~M., Gorkavyi, N.~N., Mather, J.~C.,\& Taidakova, T.~A.\
2000, ApJ, 537, L147

\bibitem[Quillen(2006)]{quillen06}
Quillen, A. C. 2006, MNRAS, 372, L14	

\bibitem[Quillen(2007)]{quillen07}
Quillen, A. C.	2007, MNRAS, 377, 1287

\bibitem[Raymond \& Barnes(2005)]{raymond05}
Raymond, S. N., \& Barnes, R.  2005, ApJ, 619, 549

\bibitem[Rieke et al.(2005)]{rieke05}
Rieke, et al. 2005, ApJ, 620, 1010

\bibitem[Strom et al.(2005)]{strom05}
Strom, Robert G., et al. 2005, Sci, 309, 1847S

\bibitem[Zuckerman \& Song(2004)]{zuckerman04}
Zuckerman, B., \& Song, I.  2004, ApJ, 603, 738	

\end{thebibliography}
\end{document}